\documentstyle[psfig,twoside,11pt]{article}

\begin{document}
\topmargin=-1.5cm
\evensidemargin=5mm
\thispagestyle{plain}
\begin{small}
\noindent{\it Astronomical and Astrophysical Transactions}, 2001

\noindent{Vol. 20, pp.393-403}
\end{small}
\vskip 1.5cm

\centerline{\Large\bf EXPERIENCE OF HIGH RESOLUTION VLBI}

\vskip 3mm

\centerline{\Large\bf IMAGING USING GENERALIZED}

\vskip 3mm

\centerline{\Large\bf MAXIMUM ENTROPY METHOD}

\vskip 1cm

\centerline{\large A.T.BAJKOVA}

\bigskip

\centerline{\it Institute of Applied Astronomy of RAS, St.Petersburg}

\bigskip

\centerline{\small\it (Received December 27, 2000)}

\bigskip

\begin{small}
{\noindent
The generalized maximum entropy method (GMEM) is a special modification
of the standard maximum entropy method (MEM) which seeks solutions in
the space of complex functions. In this work a reduced version of the GMEM
intended for reconstructing real images with positive and negative values is
used. As compared with the standard MEM, intended for the reconstruction of
only non-negative images, the GMEM allows us to obtain higher-quality images
with a much lower level of nonlinear distortions caused by errors in the data.
Here, we present the results of the GMEM imaging of 36 selected  extragalactic
radio sources with a resolution of 0.3-0.5 mas  on astrometric and geodetic
VLBI observations at 8.2 GHz, obtained with a global array in the period from
1994-1996. In VLBI mapping practice this is the first experience of
imaging with such a high resolution using maximum entropy technique.
A differential maximum entropy method intended for increasing the dynamic
range of images is demonstrated on the radio source 0059+581. In the case of
unreliable `closure' phases, completely `phaseless' methods of mapping are
recommended. Maps of two sources 0615+820 and 0642+214 are obtained using one
such method.
}

\bigskip

\noindent KEY WORDS~~~ {VLBI imaging, generalized maximum entropy
method, compact extragalactic radio sources}

\end{small}

\vskip 1cm

\noindent{\large  1~~INTRODUCTION}

\vskip 1cm

\noindent There are two main deconvolution algorithms in Very Long Baseline Interferometry
(VLBI) imaging: CLEAN and maximum entropy method (MEM). The maximum entropy
method is well known among physicists as
a very powerful tool for the nonlinear regularization of incorrect tasks.
But in
spite of its high super resolution effect, MEM is less popular in VLBI.
There are two main reasons: the high computational complexity and the nonlinear
image
distortions caused by errors in data (visibility function). But the
situation can be change fundamentally by introducing into VLBI imaging
an effective modification of the MEM called the generalized maximum entropy
method
(GMEM) which generally operates in the space of complex functions (Bajkova,
1990, 1993).
Searching for solutions of complex functions in space allows us firstly
to factor multidimensional MEM algorithms
into series of simpler
one-dimensional algorithms, which considerably decreases the computational
complexity
(Frieden and Bajkova, 1994) and, secondly, considerably decreases the nonlinear
distortions of images (Bajkova, 1995).

\newpage

\pagestyle{headings}
\makeatletter
\renewcommand{\@evenhead}{\thepage \hfil A.T.BAJKOVA \hfil}

\renewcommand{\@oddhead}{\hfil EXPERIENCE OF HIGH RESOLUTION VLBI IMAGING \hfil \thepage}

Thus, due to its generalized form MEM can become preferable in VLBI to
CLEAN, especially for imaging sources with a complicated extended structure
(we know, that MEM produces maximally smooth images, but CLEAN sharpens them).

In this paper we present the first experience of applying the GMEM to mapping
compact extragalactic sources with high resolution (0.3-0.5 mas).
We used VLBI astrometric and geodetic observations (NEOS-A program)
obtained with a global array at 8.2 GHz. Data were received from the Goddard
Space Flight Center.
Source imaging from astrometric and geodetic VLBI observations is of great
interest both
for astrometry, geodesy (geodynamics taking into account extended source
structure on mas scales to more accurate determination of
celestial/terrestrial coordinates) and astrophysics (investigation of
the short-period structure
variability of compact extragalactic radio sources).

For mapping we used the well-known `CalTech VLBI Program' package into which
we introduced the GMEM procedure as a deconvolution operation in the
selfcalibration loop.

\vskip 1cm

\noindent{\large 2~~MAPPING COMPACT EXTRAGALACTIC RADIO SOURCES}

\vskip 1cm

For VLBI source mapping a reduced version of the GMEM intended for the
reconstruction of real signals with positive and negative values was used.
The selection of the GMEM instead of the MEM as a deconvolution operation was
principally dictated by low signal-to-noise ratio typical of astrometric and
geodetic VLBI observations.

A detailed description of the practical GMEM algorithm is given in Bajkova
(1998a).
The size of maps was chosen as 256$\times$256 pixels, with an interval between two
pixels of
0.1 mas and an image domain size between 60$\times$60 and 100$\times$100 pixels, depending
in each case on structure extent and data quality. Parameter `$\alpha$'
(Bajkova, 1990, 1998a) in the GMEM algorithm, which is responsible for
separating `positive' and `negative' solutions, was chosen equal to ${10}^8$.
A numerical solution of the entropic functional was realized using
steepest-descent
method. To ensure the high stability of the numerical algorithm to
noise, an
additional regularizing term was added to the principal entropic functional
(Bajkova, 1998a).

Images of 36 selected compact extragalactic radio sources are shown in Figure~1.
Characteristics of the sources are given in Table~1. Parameters of the maps
(date of observations, list of VLBI stations, flux densities and agreement
factors) are given in Table~2.
The presented images were obtained by convolution of entropic solutions by
a circular gaussian
beam with a circle diameter on 0.2 mas. This size is enough to smooth over
image samples distant which are 0.1 mas from each other (the beam circles are
shown in bottom left corner of the maps).

The GMEM images can be compared with the corresponding CLEAN images
published in
Bajkova {\it et al.} (1996a). The comparison shows, that the  GMEM maps
are obtained with better agreement factors (nearly in 2.2 times) and are less
`lumpy'. In this connection the total flux increased on the average by 1.036
times, i.e. practically did not change. But the peak flux decreased by
1.83 times. Taking into account this and the fact that MEM and GMEM
principally seek a solution in space of smooth gaussian functions,
it is possible to conclude that the new GMEM maps are smoother than the
CLEAN maps.

\newpage

\centerline{\small{\bf Table 1.} List of sources}
\vskip 3mm
\begin{center}
\begin{small}\begin{tabular}{llllllll}
\hline
\noalign{\vskip 1mm}
\it {Source}& & \it{Alias} &\it{ Ident.} & $z$ & $m_v$ & $S_6$ & $S_{3.5}$ \\
\noalign{\vskip 1mm}
\hline
\noalign{\vskip 1mm}
0014+813&  & S5 0014+81  &         QSO  &  3.384  &  16.50 &  0.551 &  1.355 \\
0016+731&  & S5 0016+73  &         QSO  &  1.781  &  18.00 &  1.700 &  1.900 \\
0059+581&  &      -      &          -   &    -    &   -    &   -    &  2.800 \\
0202+149&  & 4C +15.05, NRAO 91    &         QSO  &    -    &  21.90 &  2.400 &  3.100 \\
0229+131&  & 4C +13.14   &         QSO  &  2.067  &  17.03 &  1.000 &  1.900 \\
0336-019&  & CTA 26      &         QSO  &  0.852  &  18.41 &  2.500 &  2.800 \\
0400+258&  &      -      &         QSO  &  2.109  &  18.00 &  1.800 &  1.400 \\
0402-362&  &      -      &         QSO  &  1.417  &  17.17 &  1.400 &   -    \\
0440+345&  &      -      &          -   &    -    &   -    &   -    &   -    \\
0458-020&  & 4C -02.19   &         QSO  &  2.286  &  19.50 &  1.900 &  3.100 \\
0528+134&  &      -      &         QSO  &  2.070  &  20.30 &  3.900 &  4.500 \\
0552+598&  & DA 193      &         QSO  &  2.365  &  18.00 &  5.400 &  5.700 \\
0615+820&  &      -      &         QSO  &  0.710  &  17.50 &  0.999 &  0.900 \\
0642+214&  & 3C 166, 4C +21.21      &         G    &  0.245  &  19.50 &  1.100 &   -    \\
0716+714&  &      -      &         LAC  &   -     &  15.50 &  1.121 &  0.600 \\
0735+178&  & OI 158      &         LAC  &  0.424  &  16.22 &  1.800 &  1.900 \\
0917+624&  & OK 630      &         QSO  &  1.446  &  19.50 &  0.996 &  1.500 \\
0955+476&  & OK 492      &         QSO  &  1.873  &  18.00 &  0.739 &  0.700 \\
1014+615&  &      -      &         BSO  &   -     &  18.10 &  0.631 &  0.571 \\
1101+384&  & Mkn 421     &         G    &  0.031  &  13.10 &  0.725 &  0.700 \\
1128+385&  & OM 346.9    &         QSO  &  1.733  &  16.00 &  0.771 &  0.900 \\
1219+285&  & W Com, ON 231       &         LAC  &  0.102  &  16.11 &  2.000 &  0.700 \\
1308+326&  & AU CVn      &         LAC  &  0.996  &  19.00 &  1.500 &  3.700 \\
1357+769&  &      -      &         QSO  &   -     &  19.00 &  0.844 &  0.600 \\
1606+106&  & 4C +10.45   &         QSO  &  1.226  &  18.50 &  1.400 &  1.600 \\
1637+574&  & OS 562      &         QSO  &  0.751  &  17.00 &  1.400 &  1.700 \\
1638+398&  & NRAO 512    &         QSO  &  1.666  &  16.50 &  1.160 &  1.500 \\
1739+522&  & 4C +51.37, OT 566     &         QSO  &  1.375  &  18.50 &  2.000 &  1.200 \\
1741-038&  &      -      &         QSO  &  1.057  &  18.60 &  3.000 &  3.000 \\
1745+624&  & 4C +62.29   &         QSO  &  3.886  &  18.70 &  0.580 &  0.480 \\
1803+784&  &      -      &         LAC  &  0.684  &  16.40 &  2.600 &  2.400 \\
1823+568&  & 4C +56.27   &         LAC  &  0.664  &  18.40 &  1.700 &  2.100 \\
2145+067&  & 4C +06.69   &         QSO  &  0.990  &  16.47 &  4.500 &  8.600 \\
2200+420&  & BL Lac, VRO 42.22.01      &         LAC  &  0.068  &  14.72 &  4.800 &  5.900 \\
2201+315&  & 4C +31.63   &         QSO  &  0.297  &  15.79 &  2.300 &  2.500 \\
2230+114&  & CTA 102, 4C +11.69     &         QSO  &  1.037  &  17.66 &  3.600 &  3.600 \\
\noalign{\vskip 1mm}
\hline
\end{tabular}\end{small}
\end{center}

\vskip 0.8cm

Improvement of agreement factors in case of GMEM shows that the space of smooth
functions for searching for source distributions is more adequate than the space
of $\delta-$ functions in which 
the CLEAN solutions are sought, i.e.
the considered sources reveal rather extended structure at the given resolution.
Note that CLEAN always gives disconnected solution as a combination of
$\delta-$ functions (in addition, CLEAN sharpens bright components and
damps weak ones), although CLEAN maps usually seem smooth. This
is because a CLEAN map is a convolution of the CLEAN solution (model) by
a smooth
gaussian `clean' beam.
And, obviously, in general a CLEAN map is not obliged to agree with the data
to
such an extent as the corresponding CLEAN model. But when we use the
GMEM we at once obtain a smooth solution which is in good agreement with
the data.

\newpage

\centerline {\small {\bf Table 2.} Parameters of maps}
\vskip 3mm
\begin{center}\begin{small}\begin{tabular}{llllllll}
\hline
\noalign{\vskip 1mm}
\it{Source} & \it{Date} & \it{Stations} &\it{Flux}  & $[Jy]$    & \it{Agr.} & \it{factor}    &       \\
         &      &         &\it{total}  & \it{peak}  & \it{ampl.}&\it{phase} &\it{total}  \\
\noalign{\vskip 1mm}
\hline
\noalign{\vskip 1mm}
0014+813  &   17.10.95 & G,W,N20,MK,NL,A,K    & 0.960  &  0.234 &  2.32 & 2.79 & 2.52   \\
0016+731  &   18.01.94 & F,W,G,K,MK,SC,N85    & 1.760  &  0.244 &  1.81 & 1.66 & 1.75   \\
0059+581  &   27.06.95 & G,W,NY,F,N85,        & 1.620  &  0.825 &  1.77 & 1.20 & 1.61   \\
0202+149  &   09.01.96 & NY,W,F,N20,G,K       & 1.310  &  0.592 &  3.15 & 1.59 & 2.77   \\
0229+131  &   18.01.94 & F,W,G,SC,N85,BR,K,MK     & 1.430  &  0.318 &  1.65 & 1.66 & 1.65   \\
0336-019  &   18.01.94 & F,W,SC,N85,BR,K,G,MK     & 2.070  &  0.737 &  2.02 & 1.92 & 1.99   \\
0400+258  &   09.01.96 & F,W,NY,N85,G,K       & 0.320  &  0.131 &  0.81 & 0.84 & 0.82   \\
0402-362  &   23.01.95 & F,H26,MA,H,S,N85,SS,MK,P        & 1.260  &  0.935 &  1.51 & 1.21 & 1.39   \\
0440+345  &   22.08.95 & K,NY,G,W,F,N85       & 0.430  &  0.131 &  1.41 & 1.95 & 1.61   \\
0458-020  &   23.01.95 & F,MA,H,N85,SC,P,     & 1.500  &  0.693 &  2.18 & 1.62 & 2.00   \\
          &            & S,MK,H26,C           &        &        &       &      &        \\
0528+134  &   01.02.95 & C,D65,MA,ME,NY,W,O,N      & 3.840  &  1.317 &  3.94 & 1.45 & 3.12   \\
0552+398  &   04.01.94 & F,N85,K,W,G,A        & 3.840  &  1.212 &  2.26 & 1.11 & 1.98   \\
0615+820  &   26.03.96 & G,K,W,NY,N20         & 0.300  &  0.079 &  1.33 & --  & --      \\
0642+214  &   14.11.95 & G,K,NY,W,F,N85       & 0.380  &  0.152 &  1.78 & -- & --       \\
0716+714  &   14.11.95 & NY,W,G,K,F,N85       & 0.180  &  0.111 &  0.96 & 1.15 & 1.02   \\
0735+178  &   18.01.94 & F,W,SC,G,BR,K,MK     & 1.540  &  0.299 &  2.59 & 2.84 & 2.68   \\
0917+624  &   05.03.96 & NY,W,N20,F,G,K       & 1.090  &  0.222 &  1.81 & 1.37 & 1.67   \\
0955+476  &   18.01.94 & G,K,W,BR,F,SC,MK     & 0.970  &  0.393 &  1.38 & 1.52 & 1.45   \\
1014+615  &   06.02.96 & G,K,N20,W,F          & 0.340  &  0.136 &  1.87 & 2.01 & 1.92   \\
1101+384  &   17.10.95 & A,G,NL,MK,K,W        & 0.350  &  0.124 &  1.43 & 2.87 & 1.14   \\
1128+385  &   31.10.95 & G,NY,K,N85,W,F       & 0.630  &  0.268 &  1.31 & 1.37 & 1.33   \\
1219+285  &   31.10.95 & G,NY,K,N85,W,F       & 0.210  &  0.067 &  0.99 & 1.00 & 0.99   \\
1308+326  &   04.10.94 & G,K,MK,W,F,SC,BR     & 3.490  &  1.249 &  3.46 & 2.96 & 3.32   \\
1357+769  &   17.10.95 & A,G,N20,W,NL,MK,K,F  & 0.590  &  0.276 &  1.52 & 2.49 & 1.96   \\
1606+106  &   18.01.94 & G,N85,K,BR,MK,F,W,SC     & 1.310  &  0.420 &  1.23 & 1.98 & 1.51   \\
1637+574  &   05.03.93 & K,N20,G,NY,W,F       & 0.850  &  0.322 &  2.15 & 1.17 & 1.87   \\
1638+398  &   08.08.95 & N85,W,G,K,F          & 0.630  &  0.276 &  1.01 & 1.76 & 1.29   \\
1739+522  &   18.01.94 & K,W,BR,MK,SC,N85,F   & 1.080  &  0.267 &  1.66 & 1.80 & 1.71   \\
1741-038  &   18.01.94 & G,K,BR,MK,N85,SC,F,W    & 2.830  &  0.898 &  1.67 & 2.75 & 2.07   \\
1745+624  &   03.10.95 & G,K,W,N20,F          & 0.200  &  0.109 &  0.61 & 0.71 & 0.64   \\
1803+784  &   18.01.94 & BR,MK,G,N85,K,SC,W,F    & 1.710  &  0.575 &  3.53 & 3.81 & 3.64   \\
1823+568  &   03.10.95 & G,W,N20,F,K          & 0.600  &  0.324 &  3.76 & 2.32 & 3.32   \\
2145+067  &   17.10.95 & A,F,G,N20,W,NL,MK,K  & 8.010  &  1.838 &  6.37 & 10.0 & 7.88   \\
2200+420  &   26.03.96 & G,K,W,NY,N20,F       & 2.810  &  0.491 &  4.38 & 2.88 & 3.97   \\
2201+315  &   18.01.94 & F,W,BR,MK,SC,N85,G,K & 1.950  &  0.306 &  1.59 & 2.56 & 1.99   \\
2230+114  &   31.10.95 & NY,W,F,N85,G,K       & 1.070  &  0.295 &  1.78 & 1.08 & 1.66   \\
\noalign{\vskip 1mm}
\hline
\end{tabular}\end{small}\end{center}

\begin{small}
\vskip 1mm
\noindent Abbreviations:
\noindent{\bf A} -- Algopark;~ {\bf BR} -- VLBA-BR;~ {\bf C} --
Crimea;~ {\bf D65} -- DSS65;~ {\bf F} -- Fortleza;~ {\bf G} --
Gilcreek;~ {\bf H} -- Hartrao;~ {\bf H26} -- Hobart26;~ {\bf K} --
Kokee;~ {\bf MA} -- Matera;~ {\bf ME} -- Medicina;~ {\bf MK} --
VLBA-MK;~ {\bf N} -- Noto;~ {\bf N20} -- NRAO20;~ {\bf N85} -- NRAO85;
~ {\bf NL} -- VLBA-NL;~ {\bf NY} -- NyAlesund20;~ {\bf O} -- Onsala;~
{\bf SC} -- VLBA-SC;~ {\bf W} -- Wettzell.
\vskip 1mm
\end{small}

\vskip 1cm

If we obtain GMEM solution with a worse agreement factor than the CLEAN
one, we should conclude that in this case GMEM is less adequate than
CLEAN and the source can be considered to be point-like source
consisting of some $\delta-$ functions. In our case, such a source is
0642+214 (in fact, its CLEAN solution has better agreement factor than
the GMEM solution (by approximately two times)).

On the basis of careful analysis of both the CLEAN and GMEM maps we can
conclude that the GMEM gave more
accurate estimates of source distributions than the CLEAN map.
In some of the sources (0400+258, 0458-020, 0917+624, 1606+106,
1823+568, 2230+114) the extended structure details are revealed to a
fuller
extent. All the images are free from the artifacts inherent in CLEAN.

To increase the dynamic range of the maps we used a differential GMEM
(see Section 3). A
more detailed comparison of CLEAN and GMEM images is given in Bajkova
(1999).

It is also necessary to note that the maps of two sources 
with unreliable `closure' phases 0615+820 and 0642+214 were obtained using
completely `phaseless' mapping
(see section 4).

\vskip 1cm

\noindent{\large 3~~A DIFFERENTIAL METHOD OF MAXIMUM ENTROPY }

\vskip 1cm

At present, the idea of differential mapping is realized most
comprehensively in the `DIFMAP' program package at the CalTech,
where CLEAN method is used as the deconvolution operation.

The method of differential mapping is based on the fundamental linear
property of
linearity of the Fourier transformation. According to this method, the bright
components of the source, which were reconstructed at the first stage,
are subtracted from the initial visibility function, the subsequent
reconstruction is performed using the residual visibility function, and
the reconstruction results are added at the final stage.

In the case of CLEAN, such a mapping method mainly influences the
convergence rate. However, if the MEM, which has pronounced nonlinear
properties, is used as the deconvolution operation, the principle of
differential
mapping can improve the reconstruction quality, particularly when the
source has bright compact components against the background of a
sufficiently weak extended base.

What is the cause of the reconstruction quality improvement? The fact is
that after the subtraction of bright components, which were
reconstructed at the first stage using the MEM, from the initial
visibility function we obtain the residual visibility function, in which
the proportion of the weak extended component becomes larger. Therefore,
we artificially decrease the dynamic interval of the map, which
corresponds to the residual visibility function, thus simplify the
image reconstruction at the second stage.

Above we discussed the advantages of the GMEM for processing real
non-negative images. With respect to differential mapping, another
advantage of GMEM originated from the fact that an image with
negative values can correspond to the residual visibility function after
subtraction of the bright component that is reconstructed at the first
stage of the algorithm from the initial visibility function. This is
possible if the bright component was reconstructed at the first stage
with an overestimated amplitude, which is quite typical of any nonlinear
method. To rule out unwanted image distortions when a non-negative
solution is sought using the data assuming the presence of negative
components, we should use the generalized method rather than the
standard one.

Simulation results of the differential GMEM are shown in Bajkova (1998b).
Here we demonstrate potential of differential GMEM for mapping the radio
source 0059+581,
which is actively used as a reference source in the astrometric and geodetic
VLBI programs. This source is of interest because it shows fast variation of
both the total flux and the structure, which consists of a compact core
and rather weak extended elements (jet). The maps that were constructed
using GMEM for a number of dates between June 1994 and
December 1995 are shown in Bajkova (1998b). Between the middle and the
end of 1994 we observed an almost linear decrease in the
total flux from $\sim$ 4 Jy to 1.5 Jy. As is obvious from Bajkova
(1998b),
we failed to obtain a map with a sufficient dynamic range detect an
extended element from June 26, 1994, when the flux was
maximal over the interval in question. Using the standard GMEM, we only
detected extended elements from October 4, 1994, when their
share with respect to the total flux became sufficient to be detected by
the selected reconstruction method. In Figure 2, for comparison, we show
maps, obtained using both the standard and differential GMEM for
three dates on which the total flux was sufficiently large and the flux
share corresponding to extended elements was insignificant. As is
obvious from the figure, the use of differential mapping allowed us to
increase the dynamic range of the maps such that the extended structures
became very distinguished.

\vskip 1cm

\noindent{\large 4~~`PHASELESS' MAPPING}

\vskip 1cm

A principle of `phaseless' mapping used for imaging two sources 0615+820 and
0642+214 (see Figure~1) is based on reconstruction of the visibility
function amplitude over the whole $UV$-plane. The image is then
reconstructed
using only the amplitude of the visibility function (the well-known `phase
problem'). For
reconstruction of the visibility amplitude GMEM has also been used in
the self-calibration loop setting all the `closure' phases to zero. For
solving the `phase problem' an effective modification of the well-known
Fienup algorithm was used (Bajkova, 1996b). `Phaseless' mapping
is recommended in the case of bad `closure' phases.

\vskip 1cm

\noindent{\large  5~~SUMMARY}

\vskip 1cm

This work is devoted to the first experience of the high resolution imaging
of compact extragalactic sources using generalized maximum entropy method
proposed by the author in earlier papers (Bajkova, 1990, 1993).
Images of a sample of 36 selected compact extragalactic radio sources
obtained with a resolution of 0.3--0.5 mas on astrometric and geodetic VLBI
observations at 8.2 GHz with a global array (NEOS-A program) between
1994 and 1996 are presented. Comparison with the corresponding CLEAN images shows
that the images obtained are smoother and characterized by better agreement
factors, revealing rather complicated structures to a fuller extent.

The choice of reconstruction method as the deconvolution procedure
in selfcalibration loop depends on the signal-to-noise ratio and {\it a
priori}
information about the source's structure. In the case of point-like
structures, which are
the majority of sources observed in geodetic programs,
the traditional CLEAN is the preferable procedure because it was invented
especially for such sources and is very fast.
In the case of complicated structures with strongly extended components,
the nonlinear information methods can be preferable. Moreover, in the
case of a low signal-to-noise ratio, which is typical for astrometric and
geodetic VLBI observations, the generalized information methods for seeking
a solution in the space of functions with positive and negative values are
preferable to the standard ones which deal only with real non-negative
distributions, because they ensure a much lower level of nonlinear distortions.

To increase the dynamic range of images we used a differential
generalized maximum entropy method. Compared with the traditional methods,
the nonlinear methods of differential mapping using the maximum entropy
method as the deconvolution operation allow us to substantially improve
the reconstruction quality of source images containing bright point
components against a background of a weak base. The maps obtained using
the differential method of maximum entropy are characterized by a higher
dynamic range. To eliminate possible nonlinear distortions in the case
of differential mapping, the generalized method of maximum entropy is
preferred to the standard method.

In the case of unreliable `closure' phases
we recommend using completely `phaseless' methods of mapping.

\vskip 1cm

\noindent{\it Acknowledgments}

\medskip

This work was supported by Federal Program `Astronomy' (Grant No 2.1.1.3.).

\bigskip

\noindent{\it References}

\medskip

{\small

Bajkova, A.T. (1990) {\it Astron. Astrophys. Trans.} {\bf 1}, 313.

Bajkova, A.T. (1993) {\it Maximum Entropy and Bayesian Methods.} Kluwer
Academic, 407.

Bajkova, A.T. (1995) {\it Izvestia vuzov. Radiofizika.} {\bf 38}, 1267
(in Russian).

Bajkova A.T. et al. (1996a)  {\it Communications of IAA RAS,} {\bf 87},
St.Petersburg (in Russian).

Bajkova, A.T. (1996b) {\it Izvestia vuzov. Radiofizika.} {\bf 39}, 472
(in Russian).

Bajkova, A.T. (1998a) {\it Trudy IPA RAN.} {\bf 3}, 287 (in Russian).

Bajkova, A.T. (1998b) {\it Izvestia vuzov. Radiofizika.} {\bf 41}, 991
(in Russian).

Bajkova, A.T. (1999) {\it Trudy IPA RAN.} {\bf 4}, 150 (in Russian).

Frieden, B.R., Bajkova, A.T. (1994) {\it Applied Optics,} {\bf 33}, 219.

\newpage

\begin{figure}
\centerline{
\psfig{figure=list1.ps,angle=-90,width=150mm}}
\centerline{
\psfig{figure=list2_.ps,angle=-90,width=150mm}
}
\noindent{\small{\bf Figure 1} Images of extragalactic sources obtained
by GMEM.
Contour levels are 1, 2, 4, 10, 20, 30, 40, 50, 60, 70, 80, 90, 99 \% of peak intensity.
}
\end{figure}

\newpage

\begin{figure}
\centerline{
\psfig{figure=list3_.ps,angle=-90,width=150mm}}
\centerline{
\psfig{figure=list4_.ps,angle=-90,width=150mm}
}
\noindent{\small{\bf Figure 1} (Continued).}
\end{figure}

\newpage

\begin{figure}
\centerline{
\psfig{figure=list5_.ps,angle=-90,width=150mm}}
\centerline{
\psfig{figure=list6.ps,angle=-90,width=150mm}
}
\noindent{\small{\bf Figure 1} (Continued).}
\end{figure}

\newpage

\begin{figure}
\centerline{
\psfig{figure=0059.ps,angle=-90,width=150mm}}
\noindent{\small{\bf Figure 2} Images of source 0059+581 obtained by
the standard GMEM (top) and the differential GMEM (bottom) for three
dates. Contour levels are 1, 2, 4, 10, 20, 30, 40, 50, 60, 70, 80, 90, 99 \% of peak
intensity.
}
\end{figure}

\end{document}